\documentstyle[12pt]{article}
\renewcommand{\theequation}{\arabic{section}.\arabic{equation}}
\title{Degeneracy and Para-supersymmetry of Dirac Hamiltonian in (2+1)- Spacetime}
\author{M.A. Jafarizadeh $^{a,b}$ \thanks{e-mail: jafarzadeh@ark.tabrizu.ac.ir} , S.K. Moayedi $^{a}$ \thanks{e-mail: moayedi@ark.tabrizu.ac.ir} \\
$^a${\small Faculty of Physics, Tabriz University, Tabriz, 51664, Iran}\\
$^b${\small Institute for Studies in Theoretical Physics and Mathematics,
Tehran, 19395-1795,Iran.}}
\begin{document}
\maketitle

\begin{abstract}

The quantum mechanics of a spin $\frac{1}{2}$ particle on a locally
spatial constant curvature part of a $(2+1)$- spacetime in the presence of 
a constant magnetic field of a magnetic monopole has been investigated. 
It has been shown that these 2-dimensional Hamiltonians have the degeneracy
group of $SL(2,c)$, and para-supersymmetry of arbitrary order or shape
invariance. Using this symmetry we have obtained its spectrum algebraically.
The Dirac's quantization condition has been obtained from the representation 
theory. Also, it is shown that the presence of angular deficit suppresses 
both the degeneracy and shape invariance.
\vspace{1cm}

{\bf Keywords: Dynamical Symmetry, Dirac Operator, Gravity, Para-}

\hspace{15mm} {\bf supersymmetry , Monopole, Shape Invariance, Spinor.}

{\bf PACs Index: 02.30. +g  or  02.90. +p.}

\end{abstract}

\vspace{9cm}

\begin{center}
\section{. INTRODUCTION}
\end{center}
Quantum theories, particularly quantum gravity in $(2+1)$- dimensions provide
us with a useful field of investigation not only for theoretical and 
mathematical issues, but also in some cases, for actual physical problems \cite{Des,Dese}.
In the past decade  many interesting physical problems in $(2+1)$-gravity 
have been solved, such as classical scattering, quantum  scattering, bound states of a
scalar and spinor point particle both in the presence and absence of a magnetic 
monopole and also magnetic vortex \cite{De,Ger,Gerb,Jaf}. 
In reference \cite{Jaf} some interesting results have been obtained in studying quantum 
scattering and bound states of a scalar charged particle in the background 
metric corresponding  to $(2+1)$- manifold both with local and global constant 
curvature in the presence of a magnetic monopole, which satisfy the coupled 
Einstein-Maxwell equations. 

Here in this article we investigate the quantum mechanics of 
a charged spin $\frac{1}{2}$ point particle on a $(2+1)$- spacetime with
spatial part of local constant curvature in the presence of 
a constant magnetic field of a magnetic monopole. We show that these 
2-dimensional Hamiltonians have the degeneracy group of $SL(2,c)$ type and 
para-supersymmetry of arbitrary order or shape
invariance. Using these symmetries we have obtained their spectra algebraically.
Also, the Dirac quantization follows naturally from the representation theory. 
In the case of local constancy of the curvature, the presence of angular 
deficit suppresses both the degeneracy and shape invariance.

The paper is organised as follows: In section II we briefly describe the $(2+1)$-
spacetime metric of reference \cite{Jaf} and assume that angular deficit is absent. 
Section III is devoted to the algorithm of the manipulation 
of the Dirac operator in these spacetimes. The Dirac operator has been given 
in terms of the generators of the $sl(2,c)$ Lie algebra, which reduces the 
familiar Dirac operator on the $S^2$ in special case \cite{Gro}.  In section IV 
we obtain the left and right invariant generators of the $SL(2,c)$ Lie group
in terms of Euler's angles \cite{Eguchi}, where after eliminating the $\psi$ 
coordinate we get the eigenspectrum of massless Dirac operator together with its 
degeneracy which is in agreement with those of reference \cite{Gro}  in special case of  
$S^2$. Also, as a special case, we obtain the massless Dirac operator in the presence of a magnetic vortex \cite{Gerb,Thaller}.
In section V we add the constant magnetic field of the magnetic monopole.  
Using again the representation of the $SL(2,c)$ Lie group we obtain the 
eigenspectrum of a charged spin $\frac{1}{2}$ particle algebraically together with its 
degeneracy group. As a special case we obtain the monopole harmonics \cite{Wu,Gros,Gross}
. Also we obtain the familiar Dirac quantization from the 
representation theory. In section VI, using the right invariant
generators and eliminating the coordinate $\psi$, we obtain the raising  and
lowering operators of the magnetic charge. Using them we show the presence of the 
para-supersymmetry of arbitrary order $p$, or equivalently, the shape invariance 
symmetry associated with the Dirac operator in the presence of a magnetic 
monopole. Finally in section VII we add the angular deficit to the background  
spacetime metric which leads to the suppression of both degeneracy and the 
shape invariance symmtery. Thus we obtain the eigenspectrum by solving the Dirac
operator by usual method which is in agreement with the result of the
reference \cite{Ger} for special case of the cone.

\begin{center}
\section{. (2+1)- Spacetime with Local Spatial Constant Curvature}
\end{center} 

In $(2+1)$ dimensions the Einstein-Hilbert action of gravity coupled to matter
and electromagnetic field, together with the cosmological term can be written as
\cite{Jaf}
\begin{equation}
S=\int d^3x \sqrt{|\det g_{_{\mu\nu}}|} \{ \frac{1}{4 \pi G}(R+2\Lambda) + 
\frac{1}{4} F_{\mu\nu}F^{\mu\nu}-{\cal L}_M \},
\end{equation}
where $\cal L_{M}$ is the matter Lagrangian corresponding to a very
massive point particle with mass $M$ located in the origin. We have rescaled G by a
factor of $4$.  
As it is shown in reference \cite{Jaf} the following $(2+1)$- spacetime metric 
corresponds to a massive point mass $M$ together with nonnegative cosmological
constant $\Lambda$ and magnetic monopole field
\begin{equation}
ds^{2}=dt^{2}-\frac{1}{2 \lambda}(d \theta^{2}+(1-GM)^2
\frac{\sin^{2}\alpha\theta}{\alpha^{2}}d\phi^{2}).
\end{equation}
where $\alpha$ and $\lambda$ satisfy the following relation
\begin{equation}
\Lambda=\alpha^2 \lambda,
\end{equation}
and $GM$ satisfies the condition $GM<1$.
The parameter $\alpha$ in equation (2.3) chooses one of the values $0,1,i$.
In the case of $\alpha=1$, $\lambda$ is a positive real number and we have
$$
0 \leq \theta < \pi,
$$
$$
0 \leq \phi < 2\pi.
$$
For $\alpha=i$, $\lambda$ is a negative real number and $(2+1)$- spacetime metric
(2.2) is Euclidean. In this case we have $0\leq \theta < \infty$ and the range
of $\phi$ is the same as $\alpha=1$ case. Finally for $\alpha=0$, $\lambda$ is
a positive real number again and $\theta$ plays the role of the radial variable
\cite{Jaf} and the range of the coordinate $\phi$ is the same as $\alpha=1,i$
case.
The magnetic monopole field corresponding to the system (2.1) is
\begin{equation}
{\cal B}=g(1-GM)\alpha\sin\alpha\theta,
\end{equation}
with
\begin{eqnarray}
g=\left \{\begin{array}{ll}
\frac{1}{2\sqrt{\pi G\lambda}}   & \mbox{if $\alpha=0$ and $1$}      \\
\frac{-1}{2\sqrt{\pi G|\lambda|}} & \mbox{if $\alpha=i$}\end{array} \right.
\end{eqnarray}
which extremizes the Einstein-Hilbert action given in (2.1). In other words they are the solution of
the Einstein-Maxwell equation which extremizes this action.
The magnetic field given in (2.4) is the magnetic field of a magnetic monople
located in the origin of the $R^3$ Euclidean space where the constant
curvature spatial part of the spacetime is embedded in it. The corresponding
magnetic potential one-form $A$ in the coordinates $\theta$ and $\phi$ is
\begin{equation}
A=-g(1-GM)\cos \alpha\theta d\phi. 
\end{equation}

In the next section after introducing the abstract Dirac operator we write
it on the spatial part of metric (2.2) in the case $M=0$.

\vspace{1cm}

\begin{center}
\section{. Dirac Operator on two dimensional spaces with global Constant 
Curvature}
\end{center} 
\setcounter{equation}{0}

The massless Dirac operator on a given d-dimensional Riemannian manifold with metric
$g_{_{\mu\nu}}$ can be written as \cite{Gro,Eguchi}:
\begin{equation}
D=-i\gamma^{a}E_{_{a}}^{\; \mu}(\partial_{_{\mu}}+\frac{1}{8}\omega_{_{\mu ab}}
[\gamma^{a},\gamma^{b}]),
\end{equation}
where $\gamma^a,s$ are the generators of the flat Clifford algebra which satisfy
the following anticommutation relation:
\begin{equation}
\{\gamma^{a},\gamma^{b}\}=2\delta^{ab} \hspace{3cm} a,b=1, \cdots ,d.
\end{equation}
Also $E_{_{a}}^{\; \mu}$ , $\omega_{_{\mu ab}}$ are d-beins and spin connections,
respectively, which satisfy the following relations:
\begin{eqnarray}
\nonumber
&&E_{_{a}}^{\; \mu}g_{_{\mu\nu}}E_{_{b}}^{\; \nu}=\delta_{_{ab}}  \\
\nonumber
&&E_{_{a}}^{\; \mu}e_{_{\mu}}^{\; b}=\delta_{_{a}}^{\; b}  \\
&&\partial_{_{\mu}}e_{_{\nu}}^{\; a}-\Gamma_{_{\mu\nu}}^{\lambda}  
e_{_{\lambda}}^{\; a}+\omega_{_{\mu ab}}e_{_{\nu}}^{\; b}=0.
\end{eqnarray}
Here in this article we are concerned with the manifolds described by metric
(2.2) and gauge potential (2.6). We also assume that $M$ vanishes in the rest of the article
except in the last section. 
Then the spatial part of the metric (2.2) reads:
\begin{equation}
ds_{_{(2)}}^{2}=\frac{1}{2 \lambda}(d \theta^{2}+\frac{\sin^{2}\alpha\theta}
{\alpha^{2}}d\phi^{2}),
\end{equation}
it is clear form the above metric that the spatial part consits of a
2-dimensional manifold of global constant curvature. Using the equations (3.3) we
obtain the following expression for the nonvanishing components of zwei-beins
and spin connections associated with metric (3.4):
\begin{eqnarray}
\nonumber
&&E_{_{1}}^{\; \theta}=\sqrt{2\lambda},  
E_{_{2}}^{\; \phi}=\sqrt{2\lambda}\frac{\alpha}{\sin\alpha\theta}  \\
&&\omega_{_{\phi 12}}=-\cos\alpha\theta.
\end{eqnarray}
By using the equation (3.1), we obtain the Dirac operator on a manifold described
by metric (3.4) as follows:
\begin{equation}
D_2=-i\sqrt{2\lambda}\gamma^{1}(\partial_{_{\theta}}+\frac{1}{2}\frac{\alpha}
{\tan\alpha\theta})-i\sqrt{2\lambda}\gamma^{2}\frac{\alpha}{\sin\alpha\theta}
\partial_{_{\phi}}.
\end{equation}
For $\alpha=1$, $D_2$ becomes the Dirac operator on the 2-dimensional sphere $S^2$
\cite{Gro}.
It is more convenient to consider the 2-dimensional manifold (3.4) as a
submanifold of the 3-dimensional manifold $M_3$ with the line element
\begin{equation}
ds_{_{(3)}}^{2}=dr^{2}+\alpha^2 r^2(d \theta^{2}+\frac{\sin^{2}\alpha\theta}
{\alpha^{2}}d\phi^{2}),
\end{equation}
which is parametrized with the following local coordinates:
\begin{eqnarray}
\nonumber
&&x_{_{1}}=r\sin\alpha\theta \cos\phi \\
\nonumber
&&x_{_{2}}=r\sin\alpha\theta \sin\phi \\
&&x_{_{3}}=r\cos\alpha\theta .
\end{eqnarray}
Now we consider $r=$ constant submanifold of $M_3$ with the following metric:
\begin{equation}
ds_{_{(2)}}^{2}=\alpha^2 r^2(d \theta^{2}+\frac{\sin^{2}\alpha\theta}
{\alpha^{2}}d\phi^{2}).
\end{equation}
For $\alpha=1$ and $r=\frac{1}{\sqrt{2\lambda}}>0$ this submanifold coincides
with the special case of $\alpha=1$ of two dimensional manifold with metric 
(3.4), while for $\alpha=0$, with the assumption of $\lim_{_{\alpha\rightarrow 
o,r\rightarrow \infty}}\alpha r=finite=\frac{1}{\sqrt{2\lambda}}$, it is 
the same as $\alpha=0$ case of metric (3.4). Finally for $\alpha=i$, with the
assumption $r=\frac{1}{\sqrt{2|\lambda|}}>0$ it changes to $\alpha=i$
case of metric (3.4).
Now we try to define the Dirac operator on the manifold $M_3$. Using the 
equations (3.3), for nonvanishing components of 3-beins and spin connections
we get
\begin{eqnarray}
\nonumber
&&E_{_{1}}^{\; \theta}=\frac{1}{\alpha r},  
E_{_{2}}^{\; \phi}=\frac{1}{r\sin\alpha\theta},  
E_{_{3}}^{\; r}=1 \\
&&\omega_{_{\theta 13}}=\alpha, \omega_{_{\phi 21}}=\cos\alpha
\theta,  \omega_{_{\phi 23}}=\sin\alpha\theta.
\end{eqnarray}
Finally, using the relations (3.1) and (3.10) the Dirac operator associated  with
metric (3.7) reads
\begin{equation}
D_3=-i\gamma^{1}\frac{1}{\alpha r}(\partial_{_{\theta}}+\frac{1}{2}\frac{\alpha}
{\tan\alpha\theta})-i\gamma^{2}\frac{1}{\alpha r}\frac{\alpha}{\sin\alpha\theta}
\partial_{_{\phi}}-i\gamma^{3}(\partial_{_{r}}+\frac{1}{r}).
\end{equation}
It is straightforward to see that the Dirac operator $D_3$ yields
\begin{eqnarray}
\nonumber
&&(-i\gamma^{3}D_3 + \frac{1}{r})\mid_{r=constant}= \\
&&-i\Gamma^{1}\frac{1}{\alpha r}(\partial_{_{\theta}}+\frac{1}{2}\frac{\alpha}
{\tan\alpha\theta})-i\Gamma^{2}\frac{1}{\alpha r}\frac{\alpha}{\sin\alpha\theta}
\partial_{_{\phi}},
\end{eqnarray}
with $\Gamma^a$ defined as  
$$
\Gamma^{a}=-i\gamma^{3}\gamma^{a},     \hspace{3cm} a=1,2
$$
which satisfy the following Clifford algebra 
\begin{equation}
\{\Gamma^{a},\Gamma^{b}\}=2\delta^{ab}.
\end{equation}
Assuming the equivalence of the metric of submanifold given in (3.9) with the 
two dimensional metric (3.4) and also replacing $\alpha r$ with $\frac{1}{
\sqrt{2\lambda}}$, we can deduce that the operator (3.12) is the same as Dirac
operator $D_2$ given in (3.6).
In terms of local coordinates (3.8) the operator $D_3$ can be written as 
\begin{equation}
D_3=-i\sigma_{_{i}}\partial_{_{i}},
\end{equation}
where $\sigma_{_{i}},s$ are Pauli matrices.
Using the identity $(\frac{\sigma_{_{i}}x_{_{i}}}{r})^2=I$, we have
\begin{equation}
D_3=(\frac{\sigma_{_{i}}x_{_{i}}}{r})^2 D_3=-i\frac{\sigma_{_{i}}x_{_{i}}}{r}
(\partial_{_{r}}+\frac{i}{r}\epsilon_{_{ijk}}\sigma_{_{i}}x_{_{j}}
\partial_{_{k}}).
\end{equation}
Now, comparing the operator (3.15) with (3.11), it follows that the operator $\gamma^3$ 
has the following form 
\begin{equation}
\gamma^{3}=\frac{\sigma_{_{i}}x_{_{i}}}{r}.
\end{equation}
Using the relations (3.12) and (3.15) together with the relation (3.11) the Dirac
operator $D_2$ over two dimensional manifold with metric (3.4) can be
represented as
\begin{equation}
D_2=\frac{1}{r}-\frac{i}{r}\epsilon_{_{ijk}}\sigma_{_{i}}x_{_{j}}
\partial_{_{k}}=\sqrt{2\lambda}(\sigma_{_{1}}I_{_{1}}+\sigma_{_{2}}I_{_{2}}
+\alpha \sigma_{_{3}}I_{_{3}}+\alpha I),
\end{equation}
where I is a $2\times 2$ identity matrix and the differential operators $I_{_{i}}$ with $i=1,2,3$ in (3.17) have
the following form:
\begin{eqnarray}
\nonumber
&& I_{_{1}}=i(\sin\phi \partial_{_{ \theta}}+   
\frac{\alpha}{\tan \alpha \theta}\cos\phi\partial_{_{\phi}}) \\
\nonumber
&& I_{_{2}}=i(-\cos\phi \partial_{_{\theta}}+  
\frac{\alpha}{\tan \alpha \theta}\sin\phi\partial_{_{\phi}}) \\
&&I_{_{3}}=-i\partial_{_{\phi}},
\end{eqnarray}
and satisfy the following $sl(2,c)$ Lie algebra
\begin{eqnarray}
\nonumber
&&[I_{_{1}},I_{_{2}}]=i\alpha^2 I_{_{3}} \\
\nonumber
&&[I_{_{2}},I_{_{3}}]=iI_{_{1}} \\
&&[I_{_{3}},I_{_{1}}]=iI_{_{2}}.
\end{eqnarray}
It is clear that for $\alpha=1$ this algebra becomes an $su(2)$ Lie algebra, 
for $\alpha=i$ we get $su(1,1)$ Lie algebra, and finally for $\alpha=0$
we get $iso(2)$ Lie algebra \cite{Jaf,Jafa}.

\begin{center}
\section{. Degeneracy group of the Dirac operator on two dimensional manifolds
with global constant curvature}
\end{center}
\setcounter{equation}{0}

In order to obtain the degeneracy group of the Dirac operator $D_2$ over two
dimensional manifold with metric (3.4) we need to know the left and right 
invariant generators of $SL(2,c)$ group which have the following form in the
Eulerean coordinates \cite{Jaf}
\begin{eqnarray}
\nonumber
&&L_{_{1}}^{(L)}=i(\sin\phi\partial_{_{\theta}}+   
\frac{\alpha}{\tan\alpha\theta}\cos\phi\partial_{_{\phi}}-
\frac{\alpha}{\sin\alpha\theta}\cos\phi\partial_{_{\psi}}) \\
\nonumber
&&L_{_{2}}^{(L)}=i(-\cos\phi\partial_{_{\theta}}+  
\frac{\alpha}{\tan\alpha\theta}\sin\phi\partial_{_{\phi}}-
\frac{\alpha}{\sin\alpha\theta}\sin\phi\partial_{_{\psi}}) \\
\nonumber
&&L_{_{3}}^{(L)}=-i\partial_{_{\phi}}, \\
\nonumber
&&L_{_{1}}^{(R)}=i(\sin\psi\partial_{_{\theta}}+   
\frac{\alpha}{\tan\alpha\theta}\cos\psi\partial_{_{\psi}}-
\frac{\alpha}{\sin\alpha\theta}\cos\psi\partial_{_{\phi}}) \\
\nonumber
&&L_{_{2}}^{(R)}=i(-\cos\psi\partial_{_{\theta}}+  
\frac{\alpha}{\tan\alpha\theta}\sin\psi\partial_{_{\psi}}-
\frac{\alpha}{\sin\alpha\theta}\sin\psi\partial_{_{\phi}}) \\
&&L_{_{3}}^{(R)}=-i\partial_{_{\psi}},
\end{eqnarray}
where $0\leq\phi <2\pi$, $0\leq\psi <4\pi$ and $0\leq\theta < \pi$ for $\alpha= 
1$, while $0\leq\theta<\infty$ when $\alpha=0,i$.
It is rather well known that both left and right invariant generators satisfy
$sl(2,c)$ Lie algebra denoted by $sl(2,c)_{_{L}}$ and  $sl(2,c)_{_{R}}$, 
respectively and also they commute with each other. That is we have 
\begin{eqnarray}
\nonumber
&&[L_{_{1}}^{(L)},L_{_{2}}^{(L)}]=i\alpha^2 L_{_{3}}^{(L)}  \hspace{15mm}
[L_{_{2}}^{(L)},L_{_{3}}^{(L)}]=iL_{_{1}}^{(L)}  \hspace{15mm}
[L_{_{3}}^{(L)},L_{_{1}}^{(L)}]=iL_{_{2}}^{(L)},  \\
\nonumber
&&[L_{_{1}}^{(R)},L_{_{2}}^{(R)}]=i\alpha^2 L_{_{3}}^{(R)}  \hspace{15mm}
[L_{_{2}}^{(R)},L_{_{3}}^{(R)}]=iL_{_{1}}^{(R)}  \hspace{15mm}
[L_{_{3}}^{(R)},L_{_{1}}^{(R)}]=iL_{_{2}}^{(R)},  \\
&&[\vec{L^{(L)}},\vec{L^{(R)}}]=0.
\end{eqnarray}
Now, using the generators (4.1) we define the following new bases:
\begin{eqnarray}
\nonumber
&&K_{_{1}}^{(L)}:=L_{_{1}}^{(L)}\otimes I+\frac{1}{2}\alpha\sigma_{_{1}} 
\hspace{12mm} 
K_{_{2}}^{(L)}:=L_{_{2}}^{(L)}\otimes I+\frac{1}{2}\alpha\sigma_{_{2}} 
\hspace{12mm} 
K_{_{3}}^{(L)}:=L_{_{3}}^{(L)}\otimes I+\frac{1}{2}\sigma_{_{3}}, \\
&&K_{_{1}}^{(R)}:=L_{_{1}}^{(R)}\otimes I 
\hspace{25mm} 
K_{_{2}}^{(R)}:=L_{_{2}}^{(R)}\otimes I 
\hspace{25mm} 
K_{_{3}}^{(R)}:=L_{_{3}}^{(R)}\otimes I.
\end{eqnarray}
Using the commutation relations (4.2) and the properties of Pauli matrices it is
rather straightforward to show that the newly defined left and right invariant
operators (4.3) also satisfy $sl(2,c)$  Lie algebra separately and commute 
with each other.
Now, the operator $F$ defined as 
\begin{equation}
F:=\sqrt{2\lambda}(\sigma_{_{1}}L_{_{1}}^{(L)}+\sigma_{_{2}}L_{_{2}}^{(L)}
+\alpha\sigma_{_{3}}L_{_{3}}^{(L)}+\alpha I),
\end{equation}
commute with all the generators given in (4.3), that is
\begin{eqnarray}
\nonumber
&&[F,\vec{K^{(L)}}]=0, \\  
&&[F,\vec{K^{(R)}}]=0.
\end{eqnarray}  
Therefore, in order to obtain the eigenspectrum of operator $F$ , we need the
set of commutative operators expressed in terms of operators (4.3),which are
$$
\{K_{_{3}}^{(R)},K_{_{3}}^{(L)},K_{_{1}}^{(L)\;2}+K_{_{2}}^{(L)\;2}+
\alpha^2 K_{_{3}}^{(L)\;2},K_{_{1}}^{(R)\;2}+K_{_{2}}^{(R)\;2}+
\alpha^2 K_{_{3}}^{(R)\;2}\}.
$$
Then, we have the following simultaneous eigenvalue equations
\begin{eqnarray}
\nonumber
&&K_{_{3}}^{(R)}\Psi=q\Psi \\
\nonumber
&&K_{_{3}}^{(L)}\Psi=m\Psi \\
\nonumber
&&(K_{_{1}}^{(R) \;2}+K_{_{2}}^{(R) \;2}+\alpha^2K_{_{3}}^{(R) \;2})\Psi=
\alpha^2 l(l+1)\Psi \\
&&(K_{_{1}}^{(L) \;2}+K_{_{2}}^{(L) \;2}+\alpha^2K_{_{3}}^{(L) \;2})\Psi=
\alpha^2 j(j+1)\Psi.
\end{eqnarray}
Obviously the operators $K_{_{3}}^{(R)}$ and $K_{_{3}}^{(L)}$ have the following
differential form 
\begin{eqnarray}
\nonumber
&&K_{_{3}}^{(R)}=\left(\begin{array}{cc}
-i\partial_{_{\psi}} & 0  \\
o & -i\partial_{_{\psi}} 
\end{array}\right),  \\
&&K_{_{3}}^{(L)}=\left(\begin{array}{cc}
-i\partial_{_{\phi}}+\frac{1}{2} & 0  \\
o & -i\partial_{_{\phi}}-\frac{1}{2} 
\end{array}\right).
\end{eqnarray}
Therefore, the two component spinor $\Psi$ reads
\begin{equation}
\Psi=\left(\begin{array}{c}
ae^{i(m-\frac{1}{2})\phi+iq\psi} f_{_{1}}(\theta) \\
be^{i(m+\frac{1}{2})\phi+iq\psi} f_{_{2}}(\theta)
\end{array}\right).
\end{equation}
In equation (4.8) $a$ and $b$ are constants and $f_{_{1}}$ and $f_{_{2}}$ are
functions of variable $\theta$.
In order the spinor $\Psi$ to become a periodic function of $\phi$ with period 
$2\pi$ , the quantum number $m$ must be a half integer number.
Now, using the left and right invariant generators (4.1) we have:
\begin{eqnarray}
\nonumber
&&L_{_{1}}^{(L) \;2}+L_{_{2}}^{(L) \;2}+\alpha^2 L_{_{3}}^{(L) \;2}=
L_{_{1}}^{(R) \;2}+L_{_{2}}^{(R) \;2}+\alpha^2 L_{_{3}}^{(R) \;2}= \\
&&-\{\partial_{_{\theta}}^{2}+
\frac{\alpha}{\tan\alpha\theta}\partial_{_{\theta}}+
\frac{\alpha^{2}}{\sin^{2} \alpha\theta}(\partial_{_{\phi}}^{2}+
\partial_{_{\psi}}^{2}-2\cos\alpha\theta\partial_{_{\phi}}\partial_{_{\psi}})\}.
\end{eqnarray}
The operator (4.9) yields the following eigenvalue equation \cite{Vilenkin}:
\begin{eqnarray}
\nonumber
&&-\{\partial_{_{\theta}}^{2}+
\frac{\alpha}{\tan\alpha\theta}\partial_{_{\theta}}+
\frac{\alpha^{2}}{\sin^{2} \alpha\theta}(\partial_{_{\phi}}^{2}+
\partial_{_{\psi}}^{2}-2\cos\alpha\theta\partial_{_{\phi}}\partial_{_{\psi}})\}
Y_{_{nq}}^{l}(\theta ,\phi ,\psi)= \\
&& \alpha^2 l(l+1)Y_{_{nq}}^{l}(\theta ,\phi ,\psi),
\end{eqnarray}
where the eigenfunction (4.10), that is $Y_{_{nq}}^{l}(\theta,\phi,\psi)$
reads
\begin{equation}
Y_{_{nq}}^{l}(\theta ,\phi ,\psi)=e^{in\phi +iq\psi}P_{_{nq}}^{l}
(\cos\alpha\theta).
\end{equation}
On the other hand, the operators 
$K_{_{1}}^{(L)\;2}+K_{_{2}}^{(L)\;2}+\alpha^2 K_{_{3}}^{(L)\;2}$ and
$K_{_{1}}^{(R)\;2}+K_{_{2}}^{(R)\;2}+\alpha^2 K_{_{3}}^{(R)\;2}$ may be
represented in the form
$$
K_{_{1}}^{(R) \;2}+K_{_{2}}^{(R) \;2}+\alpha^2K_{_{3}}^{(R) \;2}=
\left(\begin{array}{cc}
L_{_{1}}^{(R) \;2}+L_{_{2}}^{(R) \;2}+\alpha^2L_{_{3}}^{(R) \;2} & 0 \\
0 & L_{_{1}}^{(R) \;2}+L_{_{2}}^{(R) \;2}+\alpha^2L_{_{3}}^{(R) \;2} 
\end{array}\right),
$$
$$
\hspace{-100mm}K_{_{1}}^{(L) \;2}+K_{_{2}}^{(L) \;2}+\alpha^2K_{_{3}}^{(L) \;2}=
$$
\begin{equation}
\left(\begin{array}{cc}
L_{_{1}}^{(L) \;2}+L_{_{2}}^{(L) \;2}+\alpha^2L_{_{3}}^{(L) \;2}+\frac{3}{4} 
\alpha^2 +\alpha^2 L_{_{3}}^{(L)} & \alpha(L_{_{1}}^{(L)}-iL_{_{2}}^{(L)}) \\
\alpha(L_{_{1}}^{(L)}+iL_{_{2}}^{(L)}) &
L_{_{1}}^{(L) \;2}+L_{_{2}}^{(L) \;2}+\alpha^2L_{_{3}}^{(L) \;2}+\frac{3}{4} 
\alpha^2 -\alpha^2 L_{_{3}}^{(L)} 
\end{array}\right),
\end{equation}
where the operators $L_{_{1}}^{(L)}+iL_{_{2}}^{(L)}$ and
$L_{_{1}}^{(L)}-iL_{_{2}}^{(L)}$ are the raising and lowering operators of index
$n$ of eigenfunction (4.11), that is we have
$$
(L_{_{1}}^{(L)}+iL_{_{2}}^{(L)})Y_{_{nq}}^{l}(\theta ,\phi ,\psi)=
\sqrt{\alpha^2 (l+n+1)(l-n)}Y_{_{n+1q}}^{l}(\theta ,\phi ,\psi),
$$
$$
(L_{_{1}}^{(L)}-iL_{_{2}}^{(L)})Y_{_{nq}}^{l}(\theta ,\phi ,\psi)=
\sqrt{\alpha^2 (l-n+1)(l+n)}Y_{_{n-1q}}^{l}(\theta ,\phi ,\psi).
$$
Summarizing the above explanation the eigenfunction $\Psi$ of eigenvalue
equations (4.6) reads
\begin{equation}
\Psi=\Psi_{_{l,j=l\pm \frac{1}{2},m,q}}(\theta ,\phi ,\psi)=\Omega\left(\begin
{array}{c}
\pm \sqrt{l\pm m+\frac{1}{2}}Y_{_{m-\frac{1}{2}q}}^{l}(\theta ,\phi ,\psi) \\
\sqrt{l\mp m+\frac{1}{2}}Y_{_{m+\frac{1}{2}q}}^{l}(\theta ,\phi ,\psi)
\end{array}\right),
\end{equation}
where $\Omega$ is constant of normalization.
Note that in equation (4.13) $q$ takes integer values because $m$ can take
on only half integer values as said before \cite{Vilenkin}.
Now, by taking the operator $F$ of eq (4.4) to the power two we arrive at:
$$
F^2=2\lambda(K_{_{1}}^{(L) \; 2}+K_{_{2}}^{(L) \; 2}+\alpha^2K_{_{3}}^{(L) 
\; 2}+\frac{1}{4}\alpha^2 I)=2\lambda(j+\frac{1}{2})^2 \alpha^2 I.
$$
Therefore, we have the following eigenvalue equation
\begin{equation}
F\Psi_{_{l,j=l\pm \frac{1}{2},m,q}}(\theta ,\phi ,\psi)=\pm \sqrt{
2\lambda(j+\frac{1}{2})^2 \alpha^2}\Psi_{_{l,j=l\pm \frac{1}{2},m,q}}(\theta ,\phi ,\psi).
\end{equation}
Now, transfering the factor $e^{iq\psi}$ which appears in the wavefunction (4.13)  
to the left of the operator $F$ and eliminating it from both sides of
equation (4.14), we get:
\begin{equation}
F(q)\Psi_{_{l,j=l\pm \frac{1}{2},m,q}}(\theta ,\phi)=\pm \sqrt{
2\lambda(j+\frac{1}{2})^2 \alpha^2}\Psi_{_{l,j=l\pm \frac{1}{2},m,q}}(\theta ,\phi),
\end{equation}
where $F(q)$ and $\Psi_{_{l,j=l\pm \frac{1}{2},m,q}}(\theta,\phi)$ are:
\begin{eqnarray}
\nonumber
&&F(q)=\sqrt{2\lambda}\left(\begin{array}{cc}
\alpha(1-i\partial_{_{\phi}}) & e^{-i\phi}(-\partial_{_{\theta}}+i
\frac{\alpha}{\tan\alpha\theta}\partial_{_{\phi}}+q\frac{\alpha}{\sin\alpha\theta})  \\
e^{i\phi}(\partial_{_{\theta}}+i
\frac{\alpha}{\tan\alpha\theta}\partial_{_{\phi}}+q\frac{\alpha}{\sin\alpha\theta}) &
\alpha(1+i\partial_{_{\phi}})
\end{array}\right), \\
&&\Psi_{_{l,j=l\pm\frac{1}{2},m,q}}(\theta,\phi)=\Omega \left(\begin{array}{c}
\pm\sqrt{l\pm m+\frac{1}{2}}e^{i(m-\frac{1}{2})\phi}P_{_{m-\frac{1}{2} q}}
^{l}(\cos\alpha\theta) \\
\sqrt{l\mp m+\frac{1}{2}}e^{i(m+\frac{1}{2})\phi}P_{_{m+\frac{1}{2} q}}
^{l}(\cos\alpha\theta) 
\end{array}\right).
\end{eqnarray}
Now we consider the limiting case of $q\rightarrow 0$. In this limit the operator
$F(q)$ becomes the same as the operator $D_2$ in (3.17), that is we have
$$
\lim_{_{q\rightarrow o}}F(q)=D_2.
$$
Specially for $\alpha=1$ we get the Dirac operator on $S^2$ \cite{Gro}, and the
eigenfunction introduced  in (4.16) becomes the wellknown spinor 
harmonics \cite{Wu}.

There is another interesting limiting case: to let $\alpha
\rightarrow 0$ and $l\rightarrow\infty$ but $\alpha l$ to remain finite,
that is
$$                 
\lim_{_{\alpha\rightarrow 0,l\rightarrow\infty}} \alpha l=k.
$$
In this limit $\theta$ plays the role of radial coordinate and we have 
\cite{Vilenkin} 
$$
\lim_{_{\alpha\rightarrow 0,l\rightarrow\infty}}P_{_{nq}}^{l}(\cos\alpha
\theta)=J_{_{|n-q|}}(kr),
$$
where $J_{_{|n-q|}}(kr)$ is the Bessel function with index $|n-q|$.
In brief, we have 
\begin{equation}
Z_{_{k,m,q}}(r,\phi)=\lim_{_{\alpha\rightarrow 0,l\rightarrow\infty}}
\Psi_{_{l,j=l\pm\frac{1}{2},m,q}}(\theta,\phi)=\Omega^{\prime} 
\left(\begin{array}{c}
\pm e^{i(m-\frac{1}{2})\phi}J_{_{|m-\frac{1}{2}-q|}}(kr) \\
e^{i(m+\frac{1}{2})\phi}J_{_{|m+\frac{1}{2}-q|}}(kr)
\end{array}\right),
\end{equation}
where $\Omega^{\prime}$ is the new constant of normalization.
The operator $F(q)$, in the limit of $\alpha\rightarrow 0$ reads 
\begin{equation}
\lim_{_{\alpha\rightarrow 0}}F(q)=\sqrt{2\lambda}\left(\begin{array}{cc}
0 & e^{-i\phi}(-\partial_{_{r}}+\frac{i}{r}\frac{\partial}{
\partial \phi}+\frac{q}{r})  \\
e^{i\phi}(\frac{\partial}{\partial r}+\frac{i}{r}\frac{\partial}{
\partial \phi}+\frac{q}{r}) & 0
\end{array}\right).
\end{equation}
In this limit the operator $F(q)$ has the following eigenvalue 
$$
E=\pm \sqrt{2\lambda}k.
$$
The operator (4.18)  is exactly the Dirac operator of a very light spin 
$\frac{1}{2}$ particle in the presence of magnetic vortex with gauge potential   
$\vec{A}=e_{_{\phi}}\frac{q}{r}$ \cite{Thaller}. The wavefunction (4.17) is 
the eigenstate associated  with the scattering of a massless fermion from a vortex. 
it is obvious that in the case $q=0$, (4.17) and (4.18) represent the wavefunction
and the Dirac operator of a free massless fermion on two dimensional  flat space respectively.
In the next section we obtain the eigenspectrum of the Dirac operator in the
presence of the magnetic monopole (2.4) for the special case of $M=0$.

\begin{center}
\section{. The Dirac operator on a 2-dimensional manifold with global constant 
curvature in the presence of magnetic field of a magnetic monopole}
\end{center}
\setcounter{equation}{0}

The massless Dirac operator on a Riemannian manifold with metric $g_{_{\mu\nu}}$ in the presence
of gauge field $A_{_{\mu}}$ is \cite{Nakahara}
\begin{equation}
D(A)=-i\gamma^{a}E_{_{a}}^{\; \mu}(\partial_{_{\mu}}+\frac{1}{8}\omega_{_{\mu ab}}
[\gamma^{a},\gamma^{b}]+ieA_{_{\mu}}).
\end{equation}
Therefore, using the beins and spin connections given in (3.5) and considering 
the gauge potential $A=-g\cos\alpha\theta d\phi$ we obtain the following 
expression for the Dirac operator
\begin{equation}
D_2(A)=-i\sqrt{2\lambda}\gamma^{1}(\partial_{_{\theta}}+\frac{1}{2}\frac{\alpha}
{\tan\alpha\theta})-i\sqrt{2\lambda}\gamma^{2}\frac{\alpha}{\sin\alpha\theta}
(\partial_{_{\phi}}-ieg\cos\alpha\theta).
\end{equation}
Now we try to obtain the Dirac operator $D_2(A)$ given in (5.2) from the Dirac
operator on the manifold described by the metric (3.7) and by the gauge field with
connection 
\begin{eqnarray}
\nonumber
&&A_{_{r}}=0 \\
\nonumber
&&A_{_{\theta}}=0 \\
&&A_{_{\phi}}=-g\cos\alpha\theta. 
\end{eqnarray}
Using the beins and spin connections given in (3.10) and gauge field connection
(5.3), the Dirac operator on the manifold (3.7) in the presence of gauge field
(5.3) reads: 
\begin{equation}
D_3(A)=-i\gamma^{1}\frac{1}{\alpha r}(\partial_{_{\theta}}+\frac{1}{2}\frac{\alpha}
{\tan\alpha\theta})-i\gamma^{2}\frac{1}{\alpha r}\frac{\alpha}{\sin\alpha\theta}
(\partial_{_{\phi}}-ieg\cos\alpha\theta)-i\gamma^{3}(\partial_{_{r}}+\frac{1}{r}).
\end{equation}
It is straightforward to show that the following relation between the operators
$D_2(A)$ and $D_3(A)$ holds: 
\begin{equation}
D_2(A)=(-i\gamma^{3}D_3(A)+ \frac{1}{r})\mid_{r=constant}. 
\end{equation}
The gauge field (5.3) has the following form in the cartesian-like coordinates
(3.8)
\begin{equation}
A_{_{i}}=g\epsilon_{_{ij3}} \frac{x_{_{j}}x_{_{3}}}{r(x_{_{1}}^2+x_{_{2}}^2)},
\hspace{2cm} i,j=1,2,3,
\end{equation}
where it satisfies the following gauge condition
$$
\vec{r}.\vec{A}=0.
$$
In the local coordinates (3.8) together with the gauge connection (5.6), the Dirac operator
can be written as
\begin{equation}
D_3(A)=\sigma_{_{i}}(\frac{1}{i}\partial_{_{i}}+eA_{_{i}}).
\end{equation}
With further little algebra one can show that the Dirac operator given in
(5.7) takes the following form:
\begin{equation}
D_3(A)=(\frac{\sigma_{_{i}}x_{_{i}}}{r})^2D_3(A)=-i\gamma^{3}(\partial_{_{r}}-
\frac{1}{r}\vec{\sigma}.\vec{r}\times(\frac{1}{i}\vec{\nabla}+e\vec{A})).
\end{equation}
Finally using the relation (5.5) between the operators $D_2(A)$ and $D_3(A)$,
the Dirac operator on the 2-dimensional manifold (3.4) and in the presence of
magnetic field of a magnetic monopole is
\begin{equation}
D_2(A)=F(eg)-\alpha\sqrt{2\lambda}eg\gamma^{3}.
\end{equation}
Therefore, according to section 4, we have the following eigenvalue equation  
\begin{equation}
D_2(A)\Psi_{_{l,j=l\pm\frac{1}{2},m,q}}(\theta ,\phi)=\pm\sqrt{2\lambda
\alpha^2 [(j+\frac{1}{2})^2 -q^2]}\Psi_{_{l,j=l\pm\frac{1}{2},m,q}}(\theta ,\phi),
\end{equation}
where $q$ is equal to the product of electric and magnetic charge, that is
$$
q=eg.
$$
Therefore, the Dirac quantization condition follows naturally from the finite
representation of $SL(2,c)$ Lie group.
Also $j+\frac{1}{2}\geq q$ and for $j+\frac{1}{2}=q$ the operator (5.9)
becomes noninvertible.
It is clear that for $\alpha=1$, the operator $D_2(A)$ becomes the Dirac operator
on $S^2$ in the presence of magnetic field of magnetic monopole \cite{Gros,Gross}
with monopole harmonics as its eigenfunctions.

\begin{center}
\section{. Para-supersymmetry and shape invariance of Dirac equation}
\end{center}
\setcounter{equation}{0}

In this section using the left and right invariant generators introduced in 
section 4, we try  to investigate the shape invariance symmetry and
para-supersymmetry of 2-dimensional Dirac operator.
Here it is more convenient to work with bases $\{J_{_{+}}^{(R)},J_{_{-}}^{(R)},J_{_{3}}^{(R)}\}$ rather than with $\{L_{_{1}}^{(R)},L_{_{2}}^{(R)},
L_{_{3}}^{(R)}\}$ which are defined as:
\begin{eqnarray}
\nonumber
&&J_{_{\pm}}^{(R)}=L_{_{1}}^{(R)}\pm iL_{_{2}}^{(R)}=
e^{\pm i\psi}(\pm \partial_{_{\theta}}+i\frac{\alpha}{\tan\alpha\theta} 
\partial_{_{\psi}}-i\frac{\alpha}{\sin\alpha\theta} 
\partial_{_{\phi}}),  \\
&&J_{_{3}}^{(R)}=L_{_{3}}^{(R)}=-i\partial_{_{\psi}}.
\end{eqnarray}
Clearly these new bases have the following commutation relations
\begin{eqnarray}
\nonumber
&&[J_{_{+}}^{(R)},J_{_{-}}^{(R)}]=2\alpha^2 J_{_{3}}^{(R)},  \\  
&&[J_{_{3}}^{(R)},J_{_{\pm}}^{(R)}]=\pm J_{_{\pm}}^{(R)}.
\end{eqnarray}
Using the relations (4.9) and (4.10) we arrive at: 
\begin{eqnarray}
\nonumber
&&(J_{_{+}}^{(R)}\otimes I)(J_{_{-}}^{(R)}\otimes I)
\Psi_{_{l,j=l\pm \frac{1}{2},m,q}}(\theta ,\phi ,\psi)=\alpha^2(l-q+1)(l+q)
\Psi_{_{l,j=l\pm \frac{1}{2},m,q}}(\theta ,\phi ,\psi), \\
&&(J_{_{-}}^{(R)}\otimes I)(J_{_{+}}^{(R)}\otimes I)
\Psi_{_{l,j=l\pm \frac{1}{2},m,q}}(\theta ,\phi ,\psi)=\alpha^2(l+q+1)(l-q)
\Psi_{_{l,j=l\pm \frac{1}{2},m,q}}(\theta ,\phi ,\psi).
\end{eqnarray}
The above relations indicate that $J_{_{+}}^{(R)}\otimes I$ and $J_{_{-}}^{(R)}\otimes I$
are raising and lowering operators of index $q$ respectively, that is:
\begin{eqnarray}
\nonumber
&&J_{_{+}}^{(R)}\otimes I
\Psi_{_{l,j=l\pm \frac{1}{2},m,q}}(\theta ,\phi ,\psi)=\sqrt{\alpha^2
(l+q+1)(l-q)}\Psi_{_{l,j=l\pm \frac{1}{2},m,q+1}}(\theta ,\phi ,\psi), \\
&&J_{_{-}}^{(R)}\otimes I
\Psi_{_{l,j=l\pm \frac{1}{2},m,q}}(\theta ,\phi ,\psi)=\sqrt{\alpha^2
(l-q+1)(l+q)}\Psi_{_{l,j=l\pm \frac{1}{2},m,q-1}}(\theta ,\phi ,\psi).
\end{eqnarray}
Now by transfering $e^{iq\psi}$, which is only $\psi$ dependent factor in
the eigenspinor $\Psi_{_{l,j=l\pm \frac{1}{2},m,q}}(\theta ,\phi ,\psi)$, to
the left-hand sides of the lowering and raising operator in (6.4) we arrive at 
\begin{eqnarray}
\nonumber
&&J_{_{+}}^{(R)}(q)\otimes I
\Psi_{_{l,j=l\pm \frac{1}{2},m,q}}(\theta ,\phi)=\sqrt{\alpha^2
(l+q+1)(l-q)}\Psi_{_{l,j=l\pm \frac{1}{2},m,q+1}}(\theta ,\phi), \\
&&J_{_{-}}^{(R)}(q)\otimes I
\Psi_{_{l,j=l\pm \frac{1}{2},m,q}}(\theta ,\phi)=\sqrt{\alpha^2
(l-q+1)(l+q)}\Psi_{_{l,j=l\pm \frac{1}{2},m,q-1}}(\theta ,\phi),
\end{eqnarray}
where $J_{_{\pm}}^{(R)}(q)$ read:
\begin{equation}
J_{_{\pm}}^{(R)}(q)=\pm \frac{\partial}{\partial\theta}-i\frac{\alpha}{\sin
\alpha\theta}\frac{\partial}{\partial\phi}-q\frac{\alpha}{\tan\alpha\theta}. 
\end{equation}
But $J_{_{+}}^{(R)}(q)\otimes I$ and $J_{_{-}}^{(R)}(q)\otimes I$ are  still
raising and lowering operators of index $q$ of the eigenspinors
$\Psi_{_{l,j=l\pm \frac{1}{2},m,q}}(\theta ,\phi)$ and the relations (6.5)
indicate that $\Psi_{_{l,j=l\pm \frac{1}{2},m,q}}(\theta ,\phi)$ can form
the basis for a representation of para-supersymmetry of order $p$, where $p$ is an arbitrary
integer.
According to \cite{Jafa,Jafar}, the non-unitary para-supersymmetric algebra of order $p$
can be generated by parafermionic generators of order $p$, denoted by
$Q_1$ and $Q_2$ and a bosonic generator $H$, which satisfy the following
relations
\begin{eqnarray}
\nonumber
&&Q_1^pQ_2+Q_1^{p-1}Q_2Q_1+ \cdots +Q_1Q_2Q_1^{p-1}+Q_2Q_1^p=2pQ_1^{p-1}H \hspace{40mm}  (6.7a)  \\
\nonumber
&&Q_2^pQ_1+Q_2^{p-1}Q_1Q_2+ \cdots +Q_2Q_1Q_2^{p-1}+Q_1Q_2^p=2pQ_2^{p-1}H \hspace{40mm}  (6.7b) \\
\nonumber                                    
&&Q_1^{p+1}=Q_2^{p+1}=0  \hspace{114mm}    (6.7c)  \\
\nonumber                         
&&[H,Q_1]=[H,Q_2]=0.     \hspace{105mm}    (6.7d)
\end{eqnarray}
\setcounter{equation}{7}
By introducing the operators
\begin{eqnarray}
\nonumber
&&X_{_{+}}(q):=J_{_{+}}^{(R)}(q)\otimes I \\
\nonumber
&&X_{_{-}}(q):=J_{_{-}}^{(R)}(q)\otimes I   \\
&&{\cal H}_{q}:=H_{q}\otimes I,
\end{eqnarray}
we can represent the generators $Q_1$, $Q_2$ and $H$ by the following
$(p+1)\times(p+1)$ matrices of the form
\begin{eqnarray}
\nonumber
&&(Q_1)_{_{qq^{\prime}}}:=X_{_{-}}(q)\delta_{_{q+1,q^{\prime}}} \\
\nonumber
&&(Q_2)_{_{qq^{\prime}}}:=X_{_{+}}(q^{\prime}-1)\delta_{_{q,q^{\prime}+ 
1}}  \\
&&(H)_{_{qq^{\prime}}}:={\cal H}_{_{q}}\delta_{_{q,q^{\prime}}} 
\hspace{3cm}  q,q^{\prime}=1, \cdots ,p+1,
\end{eqnarray}
where each element of these matrices is a $2\times 2$ matrix.
In (6.9) we need to choose the Hamiltonians ${\cal H}, with q=1, \cdots ,p+1$ so that
the generators (6.9) satisfy the para-supersymmetric  algebraic relations
(6.7).
The generators $Q_1$, $Q_2$ and $H$, as defined in (6.9), satisfy the equation 
(6.7c), but the equations (6.7a,b) lead to the following equations:
$$
X_{_{+}}(p-2) \cdots X_{_{+}}(1)X_{_{+}}(0)X_{_{-}}(1)+ \cdots +X_{_{+}}(p-2)X_{_{-}}(p-1)X_{_{+}}(p-2)X_{_{-}}(p-3) \cdots X_{_{+}}(0)+ 
$$
\vspace{-7mm}
\renewcommand{\theequation}{\arabic{section}.\arabic{equation}{a}}
\begin{equation}
X_{_{-}}(p)X_{_{+}}(p-1)X_{_{+}}(p-2) \cdots X_{_{+}}(0)=2pX_{_{+}}(p-2)X_{_{+}}(p-3) \cdots X_{_{+}}(0){\cal H}_{_{1}}   
\end{equation}

$$
X_{_{+}}(p-1) \cdots X_{_{+}}(1)X_{_{+}}(0)X_{_{-}}(1)+X_{_{+}}(p-1) \cdots X_{_{+}}(2)X_{_{+}}(1)X_{_{-}}(2)X_{_{+}}(1)+ \cdots +   
$$
\vspace{-7mm}
\renewcommand{\theequation}{\arabic{section}.\arabic{equation}{b}}
\setcounter{equation}{9}
\begin{equation}
X_{_{+}}(p-1)X_{_{-}}(p)X_{_{+}}(p-1)X_{_{+}}(p-2) \cdots X_{_{+}}(1)=2pX_{_{+}}(p-1)X_{_{+}}(p-2) \cdots X_{_{+}}(1){\cal H}_{_{2}}   
\end{equation}

$$
X_{_{-}}(1) \cdots X_{_{-}}(p-1)X_{_{-}}(p)X_{_{+}}(p-1)+X_{_{-}}(1) \cdots X_{_{-}}(p-2)X_{_{-}}(p-1)X_{_{+}}(p-2)X_{_{-}}(p-1)+ \cdots + 
$$
\vspace{-7mm}
\renewcommand{\theequation}{\arabic{section}.\arabic{equation}{c}}
\setcounter{equation}{9}
\begin{equation}
X_{_{-}}(1)X_{_{+}}(0)X_{_{-}}(1)X_{_{-}}(2) \cdots X_{_{-}}(p-1)=2pX_{_{-}}(1)X_{_{-}}(2) \cdots X_{_{-}}(p-1){\cal H}_{_{p}}         
\end{equation}

$$
X_{_{-}}(2) \cdots X_{_{-}}(p-1)X_{_{-}}(p)X_{_{+}}(p-1)X_{_{-}}(p)+ \cdots +X_{_{-}}(2)X_{_{+}}(1)X_{_{-}}(2)X_{_{-}}(3) \cdots X_{_{-}}(p)+  
$$
\vspace{-7mm}
\renewcommand{\theequation}{\arabic{section}.\arabic{equation}{d}}
\setcounter{equation}{9}
\begin{equation}
X_{_{+}}(0)X_{_{-}}(1)X_{_{-}}(2) \cdots X_{_{-}}(p)=2pX_{_{-}}(2)X_{_{-}}(3) \cdots X_{_{-}}(p){\cal H}_{_{p+1}}.
\end{equation}
Finally equations (6.7d) imply the following equations   
\renewcommand{\theequation}{\arabic{section}.\arabic{equation}}
\begin{eqnarray}
&&{\cal H}_{_{q}}X_{_{-}}(q)=X_{_{-}}(q){\cal H}_{_{q+1}} \nonumber \\
&&{\cal H}_{_{q+1}}X_{_{+}}(q-1)=X_{_{+}}(q-1){\cal H}_{_{q}}.  
\end{eqnarray}
Now, defining the Hamiltonians ${\cal H}_{_{q}},q=1, \cdots ,p$ as:
\begin{eqnarray}
\nonumber
&&{\cal H}_{_{q}}=\frac{1}{2}X_{_{-}}(q)X_{_{+}}(q-1)+\frac{1}{2}C_{_{q}} I \hspace{1cm}  q=1,2,\cdots,p \\
&&{\cal H}_{_{p+1}}=\frac{1}{2}X_{_{+}}(p-1)X_{_{-}}(p)+\frac{1}{2}C_{_{p}} I.
\end{eqnarray}
By the definitions (6.12) relations (6.11) are satisfied  for the special case $q=p$. In order
the relations (6.11) to be satisfied for $q=1, \cdots ,p-1$, too we need
to choose the constants $C_{_{q}}$ as:
\begin{equation}
E_{_{q+1}}-E_{_{q}}=C_{_{q}}-C_{_{q+1}},
\end{equation}
where
\begin{equation}
E_{_{q}}:=\alpha^2[l(l+1)-q(q-1)].
\end{equation}
To obtain relation (6.13) we have used the following shape
invariance property between the operators $X_{_{\pm}}(q)$
\begin{equation}
X_{_{+}}(q-1)X_{_{-}}(q)-X_{_{-}}(q+1)X_{_{+}}(q)=E_{_q}-E_{_{q+1}}.
\end{equation}
Substituting (6.12) in formula (6.10a), and also using the shape invariance  
property (6.15) we obtain 
\begin{equation}          
C_{_1}=\frac{1}{p}[(1-p)E_{_{1}}+E_{_{2}}+E_{_{3}}+ \cdots +E_{_p}].
\end{equation}
Finally combining (6.13) with (6.16) we obtain
\begin{equation}
C_{_{q}}=\frac{1}{p}\sum_{q^{\prime}=1}^{p}E_{_{q^{\prime}}}-E_{_{q}}.
\end{equation}
From (6.17) we can see that the following relation 
among the constants $C_{_{q}}$ holds
$$
C_{_{1}}+C_{_{2}}+ \cdots +C_{_{p}}=0.
$$
Using the substitution (6.12) and the shape invariance property (6.15) together
with the constants $C_{_{q}}$ given in (6.17) one can straightforwardly show
that all the equations (6.10b,c,d) are satisfied, too. Also using the result 
given in (6.17) and the shape invariance relation (6.15) it follows that
the Hamiltonians ${\cal H}_{_{q}}$ are isospectral and we have 
\begin{equation}
{\cal H}_{_{q}}\Psi_{_{l,j=l\pm \frac{1}{2},m,q-1}}(\theta ,\phi)=E
\Psi_{_{l,j=l\pm \frac{1}{2},m,q-1}}(\theta ,\phi),  \hspace{3cm} q=1, \cdots ,p+1. 
\end{equation}
with
\begin{equation}
E=\frac{1}{2p}\sum_{q=1}^{p}E_{_{q}}.
\end{equation}
Substituting $E_{_{q}}$ in (6.19) and by using the relation (6.14) we get 
\begin{equation}
E=\frac{1}{6}\alpha^{2}[3l(l+1)+1-p^{2}].
\end{equation}
In a similar manner by substituting (6.14) in (6.17) we have 
\begin{equation}
C_{_{q}}=\frac{1}{3}\alpha^{2}[3q(q-1)+1-p^{2}].
\end{equation}
Substituting the constants $C_{_{q}}$ in (6.21), and also using the relations
(6.6) and (6.8) after the substituting (6.12) we obtain the explicit differential
form of the Hamiltonian ${\cal H}_{_{q}}$ as 
\begin{equation}
{\cal H}_{_{q}}=-\frac{1}{2}[\frac{\partial^{2}}
{\partial\theta^{2}}+\frac{\alpha}{\tan\alpha\theta}\frac{\partial}{\partial\theta}
+\frac{\alpha^{2}}{\sin^{2}\alpha\theta}\frac{\partial^{2}}{\partial\phi^{2}}
-\frac{2i(q-1)\alpha^{2}}{\sin\alpha\theta \tan\alpha\theta}\frac{\partial}{\partial\phi} 
-\frac{(q-1)^2\alpha^{2}}{\sin^{2}\alpha\theta}+\frac{1}{3}\alpha^2(p^2-1)]\otimes I.  
\end{equation}
The bases of the representation of para-supersymmetric algebra of order $p$
can be represented by column matrix with $(p+1)$ row, that is
\begin{equation}
(\Psi_{_{l,j=l\pm \frac{1}{2},m}}(\theta ,\phi))_{_{q}}:=
\Psi_{_{l,j=l\pm \frac{1}{2},m,q}}(\theta ,\phi),  \hspace{3cm}
q=0,1, \cdots ,p, 
\end{equation}
where $\Psi_{_{l,j=l\pm \frac{1}{2},m}}(\theta ,\phi)$ is the eigenvector of paraboson operator $H$ with eigenvalue $E$, that is
\begin{equation}
H\Psi_{_{l,j=l\pm \frac{1}{2},m}}(\theta ,\phi)=E
\Psi_{_{l,j=l\pm \frac{1}{2},m}}(\theta ,\phi).
\end{equation}
It follows rather trivially from the commutation relations (6.7d) and also from the
relations $(6.7c)$  that $Q_1^{q}\Psi_{_{l,j=l\pm\frac{1}{2},m}}(\theta,\phi)$
and $Q_2^{q}\Psi_{_{l,j=l\pm\frac{1}{2},m}}(\theta,\phi)$ for $q=1, \cdots ,p$
are eigenstates of the bosonic generator $H$ with the corresponding eigenvalue
$E$. Hence, it follows that 
\begin{equation}
\Psi_{_{l,j=l\pm \frac{1}{2},m,q}}(\theta ,\phi)=
\frac{X_{_-}(q+1)}{\sqrt{E_{_{q+1}}}}\frac{X_{_-}(q+2)}{\sqrt{E_{_{q+2}}}} 
\cdots \frac{X_{_-}(q^{\prime}+q)}{\sqrt{E_{_{q^{\prime}+q}}}}
\Psi_{_{l,j=l\pm \frac{1}{2},m,q^{\prime}+q}}(\theta ,\phi), 
\hspace{10mm} q=0,1, \cdots ,p-q^{\prime}.
\end{equation}
The representation of para-supersymmetry algebra of order $p$ is valid even
for the special case of $\alpha=0$. Using the relation (4.17) in the limiting cases
of $\alpha\rightarrow 0$ and $l\rightarrow\infty$ we arrive at the following 
shape invariance relations
\begin{eqnarray}
\nonumber
&&(\frac{\partial}{\partial r}-\frac{i}{r}\frac{\partial}
{\partial \phi}-\frac{q}{r})\otimes I Z_{_{k,m,q}}(r,\phi)=k
Z_{_{k,m,q+1}}(r,\phi) \\
&&(-\frac{\partial}{\partial r}-\frac{i}{r}\frac{\partial}
{\partial \phi}-\frac{q}{r})\otimes I Z_{_{k,m,q}}(r,\phi)=k
Z_{_{k,m,q-1}}(r,\phi).
\end{eqnarray}
Note that in the limiting case of $\alpha=0$ there is neither any
highest nor lowest state in the realization of para-supersymmetry, since there
is no first order differential operator which can kill any of the Bessel
functions. In this case we can have a para-supersymmetry of infinite order where
the Bessel functions form its bases.

\begin{center}
\section{. Solution of Dirac equation on 3-dimensional manifolds with a local
constant curvature}
\end{center}
\setcounter{equation}{0}

For nonvanishing $M$, the spatial part of the metric given in (2.2) describes 
a 2-dimensional manifold with local constant curvature. In this section we try
to solve the Dirac equation of a point mass ${\cal M}$ and charge $e$ on the
manifold with metric (2.2), in the presence of magnetic field with connection 
(2.6). The angular deficit due to the presence of a very heavy point mass $M$
destroys the global constancy of the curvature, hence we lose both the
degeneracy and para-supersymmetry. Therefore, we can't solve the Dirac equation
by the algebraic methods anylonger. Thus we have to solve it by an ordinary method
of solution of coupled first order differential equations.
In a flat $(2+1)$- spacetime one can represent the Dirac $\gamma$ matrices as
\cite{Ger,Gerb}
$$
\gamma^0=\sigma_{_{3}}
$$
$$
\gamma^1=i\sigma_{_{2}}
$$
\begin{equation}
\gamma^2=-i\sigma_{_{1}},
\end{equation}
where $\gamma^{a}, \; a=0,1,2$ close the Clifford algebra as 
\begin{equation}
\{\gamma^a ,\gamma^b\}=2\eta^{ab}  \hspace{3cm} a,b=0,1,2.
\end{equation}
The Minkowski metric $\eta_{_{ab}}$ has the following signature
$$
\eta_{_{ab}}=diag(+,-,-).
$$
One can write the metric (2.2) in the following form
\begin{equation}
ds^{2}=dt^{2}-\rho^2(\beta^{-2}d \theta^{2}+
\frac{\sin^{2}\alpha\theta}{\alpha^{2}}d\phi^{2}),
\end{equation}
where $\rho^2$ and $\beta$ are defined as:
$$
\beta=1-GM
$$
$$
\rho^2=\frac{(1-GM)^2}{2\lambda}.
$$
Since $\lambda$ is negative when $\alpha=i$ we have $\rho^2 <0$, hence the
metric (7.3) is Euclidean while for other values of $\alpha$ it is Minkowskian.
We can choose the three-beins associated  with the metric (7.3) as:
\begin{equation}
e_{_{\mu}}^{\; a}=\left(\begin{array}{ccc}
1 & 0 & 0  \\
0 & \rho\frac{1}{\beta}\cos\phi & \rho\frac{1}{\beta}\sin\phi   \\ 
0 & -\rho\frac{\sin\alpha\theta}{\alpha}\sin\phi & \rho\frac{\sin\alpha\theta}{\alpha}\cos\phi
\end{array}\right).
\end{equation}
With the above choice of 3-beins 
we can compare our results in a special case with  those of reference \cite{Ger}. The inverse of matrix (7.4) is:
\begin{equation}
E_{_{a}}^{\; \mu}=\left(\begin{array}{ccc}
1 & 0 & 0  \\
0 & \frac{\beta}{\rho}\cos\phi & -\frac{1}{\rho}\frac{\alpha}{\sin\alpha\theta}\sin\phi \\ 
0 & \frac{\beta}{\rho}\sin\phi & \frac{1}{\rho}\frac{\alpha}{\sin\alpha\theta}\cos\phi 
\end{array}\right).
\end{equation}
According to reference \cite{Ger}, the Dirac equation in $(2+1)$- spacetime for a fermion with mass 
${\cal M}$ and electric charge $e$, in the presence of a gauge field with gauge
connection $A_{_{\mu}}$ is: 
\begin{equation}
[i\gamma^aE_{_{a}}^{\; \mu}(\partial_{_{\mu}}-\frac{i}{2}\omega_{_{\mu}}^{\; b}\gamma_{_{b}}+
ieA_{_{\mu}})-{
\cal M}]\Psi(t,\theta ,\phi)=0,
\end{equation}
where $\omega_{_{\mu}}^{\; a}$ is given by
\begin{equation}
\epsilon^{\kappa\mu\nu}\partial_{_{\mu}}e_{_{\nu}}^{\; a}=
\epsilon^{\kappa\mu\nu}\epsilon^{a}_{_{bc}}\omega_{_{\mu}}^{\; b}e_{_{\nu}}^{\; c}.
\end{equation}
Hence
\begin{equation}
\omega_{_{\mu}}^{\; a}=\left(\begin{array}{ccc}
0 & 0 & 0  \\
0 & 0 & 0  \\
\beta\cos\alpha\theta-1 & 0 & 0
\end{array}\right).
\end{equation}
Therefore, the Dirac equation (7.6) can be written as
\begin{equation}
\{i[\gamma^0 \partial_{_{t}}+\frac{1}{\rho}\gamma^{\theta}(\beta
\partial_{_{\theta}}-\frac{\alpha}{2\sin\alpha\theta}(1-\beta\cos\alpha\theta))
+\frac{1}{\rho}\frac{\alpha}{\sin\alpha\theta}\gamma^{\phi}(\partial_{_{\phi}}+i
eA_{_{\phi}})]-{\cal M}\}\Psi(t,\theta ,\phi)=0,
\end{equation}
where $\gamma^{\theta}$ and $\gamma^{\phi}$ are defined as:
$$
\gamma^{\theta}=\cos\phi\gamma^{1}+\sin\phi\gamma^{2}
$$
$$
\gamma^{\phi}=-\sin\phi\gamma^{1}+\cos\phi\gamma^{2}.
$$
In the limiting case of $\alpha\rightarrow 0$, the coordinate $\theta$ becomes 
similar to a radial coordinate $r$ and the Dirac equation (7.9) in the absence
of gauge field $A_{_{\phi}}$ becomes exactly the Dirac equation associated
with a massive fermion on a $(2+1)$- spacetime dimension with conical spatial
part \cite{Ger}.
Now let the Dirac spinor has the following time dependence 
$$                                                                
\Psi(t,\theta ,\phi)=e^{-iEt}\Psi(\theta ,\phi),
$$
together with the following $\phi$ dependence
\begin{equation}
\Psi(\theta,\phi)=\left(\begin{array}{c}
e^{i(m-\frac{1}{2})\phi}f_{_{1}}(\theta) \\
e^{i(m+\frac{1}{2})\phi}f_{_{2}}(\theta) 
\end{array}\right),
\end{equation}
where $m$ is a half integer number. One can show that the functions $f_{_{1}}(\theta)$
and $f_{_{2}}(\theta)$ satisfy the following differential equations
$$
\{(1-z^2)\frac{d^2}{dz^2}-2z\frac{d}{dz}-\frac{1}{1-z^2}((\frac{m}{\beta})^2+
(eg\pm\frac{1}{2})^2-2\frac{m}{\beta}(eg\pm\frac{1}{2})z)\}f_{_{1,2}}(z)=
$$
\begin{equation}
-(\frac{R^2}{\alpha^2 \beta^2}(E^2-{\cal M}^2)+(e^2g^2-\frac{1}{4}))f_{_{1,2}}(z),
\end{equation}
with $z$ defined as
$$
z=\cos\alpha\theta.
$$
Now defining 
$$
\frac{\rho^2}{\alpha^2 \beta^2}(E^2-{\cal M}^2)+(e^2g^2-\frac{1}{4})=c(c+1),
$$
with $c$ as a real number, one can give the solutions of equations (7.11) in 
terms of hypergeometric functions \cite{Grad}.
In the limiting cases  of $\alpha,e\rightarrow o$ and $c\rightarrow\infty$
such that the product $\alpha c$ remains constant we obtain \cite{Grad}
\begin{equation}
\lim_{_{\alpha\rightarrow 0,c\rightarrow \infty,e\rightarrow 0}}f_{_{1,2}}
(\theta)=J_{_{|\frac{m}{\beta}\mp\frac{1}{2}|}}(\alpha cr).
\end{equation}
Writing the half integer number $m$ as
$$
m=n+\frac{1}{2},
$$
with $n$ as an arbitrary integer the Dirac equation (7.9) takes the following solution 
in the above mentioned limit
\begin{equation}
\Psi(t,\theta,\phi)=e^{-iEt}\exp{i(n+\frac{1}{2}-\frac{1}{2}\sigma_{_{3}})\phi}
\left ( \begin{array}{c}
AJ_{_{|\frac{1}{\beta}(n+\frac{1-\beta}{2})|}}(\alpha cr) \\
BJ_{_{|\frac{1}{\beta}(n+\frac{1-\beta}{2})+1|}}(\alpha cr) 
\end{array}\right ),
\end{equation}
where $A$ and $B$ are arbitrary constants.
The solution (7.13) is eigenstate of Dirac Hamiltonian associated with a fermion
on $(2+1)$ spacetime with conical spatial part \cite{Ger} with corresponding
eigenvalues:
\begin{equation}
E=\pm\sqrt{{\cal M}^2+\frac{1}{\rho^2}\beta^2\alpha^2 c^2}.
\end{equation}

\end{document}